\documentclass[%
 reprint,
 amsmath,amssymb,
 aps,
]{revtex4-2}

\usepackage{graphicx}
\usepackage{dcolumn}
\usepackage{bm}
\usepackage{xcolor}
\usepackage{subfigure}
\usepackage{amsfonts,amsmath,amssymb}
\usepackage{comment}
\usepackage{mathrsfs}

\newcommand{\btheta}{\mathbf{\theta}}
\newcommand{\phiNN}{\phi_{\btheta}}
\newcommand{\bfr}{\mathbf{r}}

\begin{document}

\preprint{APS/123-QED}

\title{
Effective Many-body Interactions in Reduced-Dimensionality Spaces Through Neural Network Models}

\author{Senwei Liang}
\email{SenweiLiang@lbl.gov}
\affiliation{%
 Applied Mathematics and Computational Research Division\\
 Lawrence Berkeley National Laboratory 
}%
\author{Karol Kowalski}%
\email{karol.kowalski@pnnl.gov}
\affiliation{%
Physical Sciences Division,\ Pacific Northwest National Laboratory, Richland, Washington 99354, USA 
}%


\author{Chao Yang}
\email{CYang@lbl.gov}
\affiliation{%
 Applied Mathematics\ and Computational Research Division\\
 Lawrence Berkeley National Laboratory 
}%

\author{Nicholas P. Bauman}
\email{nicholas.bauman@pnnl.gov}
\affiliation{%
Physical Sciences Division, Pacific Northwest National Laboratory, Richland, Washington 99354, USA 
}%


\date{\today}

\begin{abstract}
Accurately describing properties of challenging problems in physical sciences often requires complex mathematical models that are unmanageable to tackle head-on. Therefore, developing reduced-dimensionality representations that encapsulate complex correlation effects in many-body systems is crucial to advance the understanding of these complicated problems. 
However, a numerical evaluation of these predictive models can still be associated with a significant computational overhead. To address this challenge,  in this paper, we discuss a combined framework that integrates recent advances in the development of active-space representations of coupled cluster (CC) downfolded Hamiltonians with neural network approaches. 
The primary objective of this effort is to train neural networks to eliminate the computationally expensive steps required for evaluating hundreds or thousands of Hugenholtz diagrams, which correspond to multidimensional tensor contractions necessary for evaluating a many-body form of downfolded/effective Hamiltonians. 
Using small molecular systems (the H$_2$O and HF molecules) as examples, we demonstrate that training neural networks employing effective Hamiltonians for a few nuclear geometries of molecules can accurately interpolate/extrapolate their forms to other geometrical configurations characterized by different intensities of correlation effects. 
We also discuss differences between effective interactions that define CC downfolded Hamiltonians with those of bare Hamiltonians defined by Coulomb interactions in the active spaces.
\end{abstract}

\maketitle


\section{Introduction}

The development of methodologies and approximations that effectively capture complex correlation effects is crucial in computational chemistry and physics for comprehending the diverse properties of quantum systems.
Various representations of quantum mechanics, such as wave function,\cite{jimenez2015cluster,braunscheidel2024accurate,gray2024hyperoptimized}
density matrix, Green's function, and electron density-based formulations have been employed over the last few decades to enable these approaches. For example, the Density Functional Theory (DFT) has evolved into the driving engine for numerous electronic structure simulations in chemistry and materials sciences, aiming to approximate exchange-correlation potential. On the other end of the spectrum, a hierarchical family of approximations provided by a broad class of coupled cluster (CC) formalisms offers a highly accurate characterization of ground- and excited-state processes, converging in the asymptotic limit to the exact theory. 
In many instances, formulations used in the electronic structure theory provide a universal computational framework that works across length, energy, and time scales, as demonstrated by the universality of the CC formalism used in chemistry, materials \cite{hirata2004coupled,booth2013towards,zhang2019coupled,liao2021towards,wang2020excitons}, 
and nuclear structure theory simulations
\cite{dean2004coupled,kowalski2004coupled,hagen2014coupled,hu2022ab}. 
However, the main challenge that precludes the widespread utilization of these methods is the polynomial scaling with the system size, which significantly narrows the corresponding area of applications, particularly in the context of high-accuracy methodologies. 

To address this challenge, several approaches to handle numerical overhead have been pursued, including reducing numerical costs of methodologies and embracing high-performance conventional computing, as well as attempting to utilize quickly evolving quantum computing technologies. While high-performance computing (HPC) is currently a necessary component of nearly all computational packages used in simulations, the best effect is achieved when HPC technologies are integrated with the theoretical frameworks to either reduce the scaling or the dimensionality of the problem.
Reduced-scaling formulations, commonly referred to as local formulations, enforce sparsity conditions on the dense representations of equations
\cite{riplinger2013efficient,riplinger2013natural,pinski2015sparse,riplinger2016sparse}. 
On the other hand, reduced-dimensionality approaches involve representing the problem in reduced-dimensionality subspace(s), also known as active space(s). 
In general, reduced-dimensionality representations provide an alternative means of incorporating sparsity compared to typical reduced-scaling approaches. 

In recent years, significant inroads have been made in developing downfolded Hamiltonians for quantum chemistry and quantum computing applications. These Hamiltonians are designed to accurately reproduce the system's ground-state energy in active spaces by integrating external correlation effects in the many-body form of the Hamiltonians. The CC downfolding approach results from the exponential parametrization (ansatz) for the ground-state wave function in the single-reference parametrization. It can be considered a many-body embedding or renormalization procedure when applying multi-step downfolding procedures. These formulations provide a rigorous theoretical and computational framework for calculating effective Hamiltonians, promising prospects for further advancements in quantum chemistry, quantum embedding theory, and quantum computing.\cite{kowalski2018properties,shee2024static,huang2023leveraging,kowalski2023quantum}

When it comes to utilizing CC downfolding methods in practical applications, the most significant challenge is the numerical effort required to evaluate many tensor contractions associated with the Hugenholtz diagrams that define the many-body structure of an effective Hamiltonian. 
Even HPC technologies may only provide a viable solution within a particular system size limit, and therefore a computational/evaluation algorithm change is needed for larger and more complex applications. One potential solution to these issues is using neural networks (NN). This choice of computational methodology is motivated by recent applications of deep NNs (DNN) to solve many-body Schr\"odinger equation and identify corresponding wave function
\cite{carleo2017solving,xia2018quantum,pfau2020ab,hermann2020deep} (see also 
 Ref.\cite{bassi2024learning} for the related effort to solve integro-differential equations). Our approach is driven by a different paradigm closely related to the ability of downfolded Hamiltonians to provide highly accurate estimates of the total ground-state energy. Instead of attempting to determine wave functions, which generally require evaluating a massive number of parameters and properly accounting for the anti-symmetry for Fermionic systems, we use the DNN to evaluate a limited set of interactions contained within small-dimensionality active spaces. Even though downfolded Hamiltonians generally contain interactions beyond pairwise interactions, the number of these interactions (using tensors and second-quantized representations to represent them) is significantly smaller than the number of wave function parameters for general dense systems typically encountered in quantum chemistry. 

The proposed framework can be regarded as an instance of physics-informed neural networks (PINNs) \cite{karniadakis2021physics}, which leverage the fundamental principles of the lower-dimensional representations of quantum systems to design machine learning (ML) models that take into account special structures of interactions in diverse physical scenarios.

In this paper, for the first time, we put forward a DNN framework (referred to as the VNet) to evaluate effective interactions in correlated systems. To assess the feasibility of the VNet, we focus entirely on molecular benchmark systems (the H$_2$O and HF molecules) and two-body effective interactions. The VNet DNN takes four indices of the tensor representing two-body interactions of the effective Hamiltonians in the active space, as well as the geometry of the molecule, as the input, and predicts the value of the corresponding tensor element as the output. We assume that $P$, $Q$, $R$, and $S$ spin-orbital indices designate active spin-orbitals. The output is used in a loss function that measures the discrepancy between the predicted and the true value of the tensor elements $k^{PQ}_{RS}$ contained in a training dataset that consists of easily obtainable elements of the active-space effective interactions associated with the downfolded Hamiltonians for a few selected geometries. Also included in the training dataset are two-body bare interactions $v^{PQ}_{RS}$ associated with the original Hamiltonian in the same active space.  
The DNN parameters in VNet are optimized in a batched training procedure that minimizes the loss function in an iterative scheme. The training is carried out in two steps. In the first step, several bare interaction tensors contained in the training dataset are used to initialize the VNet parameters.  These parameters are finetuned in a subsequent step in which selected effective interaction tensors associated with a few selected geometries are used to finetune the VNet parameters.
An exciting feature of VNet is that the DNN optimization cost (excluding the formation cost of training data - downfolded Hamiltonians) depends only on the number of active orbitals.

The paper is organized as follows. In section~\ref{sec:ccdownfold}, we briefly describe the coupled cluster downfolding theory that can be used to derive an effective downfolded Hamiltonian within an active space.
The second quantized representation of the downfolded effective Hamiltonian is given in section~\ref{sec:methods}. The main ideas of using machine learning ML techniques to predict effective interactions for different molecular geometries using a carefully designed and physics-informed neural network trained with bare and effective interactions at a few selected molecular geometries are presented in section~\ref{sec:ML}. We demonstrate the effectiveness of the ML approach in section~\ref{sec:results} using the H$_2$O and HF molecules in different geometric configurations as test cases. We also show that the ML approach not only allows us to predict effective interactions but also provides information that can possibly be used to characterize the nature of screening effects in the effective interactions. Some main conclusions and future directions are discussed in section~\ref{sec:conclusion}.


%

\section{Coupled cluster downfolding}
\label{sec:ccdownfold}

In contrast to the numerous effective Hamiltonian formulations discussed in physics and chemistry, the CC downfolding discussed in Refs.\cite{kowalski2018properties,bauman2019downfolding} is a consequence of the ground-state wave function's single-reference exponential representation (ansatz).
The existing two classes of CC downfolding, Hermitian and non-Hermitian CC downfolding procedures,  utilize the standard single-reference CC (SR-CC) and double unitary CC (DUCC) representations of the ground-state wave functions where the ansatz can be naturally factorized into products of exponential ansatzen defined by internal and external cluster operators generating correlation effects inside/outside the active spaces, respectively. Below, we provide a brief review of these formulations. 

\subsection{Non-Hermitian CC Downfolding}
The non-Hermitian CC downfolding is a consequence of the SR-CC exponential ansatz, where the ground-state wave function $|\Psi\rangle$ in exact or approximate representations is expressed as
\begin{equation}
|\Psi\rangle = e^T |\Phi\rangle \;,
\label{eq1}
\end{equation}
where $T$ is the cluster operator and $|\Phi\rangle$ designates the so-called reference function, usually chosen as a Hartree--Fock (HF) determinant. We also introduce the notion of 
the complete active space (CAS), which is generated by the sets of active occupied ($R$) and virtual ($S$) orbitals, where all excited Slater determinants spanning CAS
can be generated as the action of elements of the commutative sub-algebra $\mathfrak{h}\equiv \mathfrak{g}^{(N)}(R,S)$ 
(expressed in the particle-hole representation with respect to the reference Slater determinant $|\Phi\rangle$) on the reference function $|\Phi\rangle$. The sub-algebra $\mathfrak{h}$ in a natural
induces the partitioning of the cluster operator $T$ into internal, $T_{\rm int}(\mathfrak{h})$,  producing excited configurations within CAS when acting on $|\Phi\rangle$, and external, $T_{\rm ext}(\mathfrak{h})$, producing excited configurations in the orthogonal complement of CAS when acting on $|\Phi\rangle$, parts, i.e.,
\begin{equation}
T = T_{\rm int}(\mathfrak{h})+T_{\rm ext}(\mathfrak{h})
\label{eq2}
\end{equation}
and 
\begin{equation}
|\Psi\rangle = e^{T_{\rm ext}(\mathfrak{h})} 
e^{T_{\rm int}(\mathfrak{h})}
|\Phi\rangle \;.
\label{eq2a}
\end{equation}
If the $T_{\rm int}(\mathfrak{h})$ operator generates all possible excited determinants when acting on $|\Phi\rangle$ we call the $\mathfrak{h}$ sub-algebra a {\it sub-system embedding sub-algebra} (SES) for a given representation of the cluster operator $T$. 
 
 The partitioning of the cluster operator into internal and external parts has been initially introduced in the context of the state-selective CC formalism in Refs.\cite{pnl93,piecuch1994state,adamowicz1998state,piecuch_molphys} and was invoked as a selection mechanism for high-rank cluster amplitudes. An analogous representation of the cluster operator was used to develop various tailored CC methods \cite{kinoshita2005coupled,veis2016coupled}.

Employing the idea of SESs one can demonstrate that the CC can be obtained in an alternative way to the standard text-book formula 
\begin{equation}
E = \langle\Phi|e^{-T}He^T|\Phi\rangle, 
\label{eq3}
\end{equation}
namely as an eigenvalue of effective or downfolded  Hamiltonian $H^{\rm D}(\mathfrak{h})$  in the $\mathfrak{h}$-generated CAS
\begin{equation}
    H^{\rm D}(\mathfrak{h}) e^{T_{\rm int}(\mathfrak{h})} |\Phi\rangle = E e^{T_{\rm int}(\mathfrak{h})} |\Phi\rangle \;,
    \label{eq4}
\end{equation}
where 
\begin{equation}
    H^{\rm D}(\mathfrak{h}) = (P+Q_{\rm int}(\mathfrak{h}))
    e^{-T_{\rm ext}(\mathfrak{h})} H
    e^{T_{\rm ext}(\mathfrak{h})}
    (P+Q_{\rm int}(\mathfrak{h})) \;,
    \label{eq5}
\end{equation}
and $P$ and $Q_{\rm int}(\mathfrak{h})$ refer to the projection operator onto the reference function and all excited Slater determinants belonging to $\mathfrak{h}$-generated CAS. The standard energy expression (\ref{eq3}) is a special case of (\ref{eq5}) for $R$ and $S$ being empty sets (in such a case the CAS is generated by a reference function $|\Phi\rangle$).

\begin{widetext}
\begin{figure*}
\includegraphics[scale=0.43]{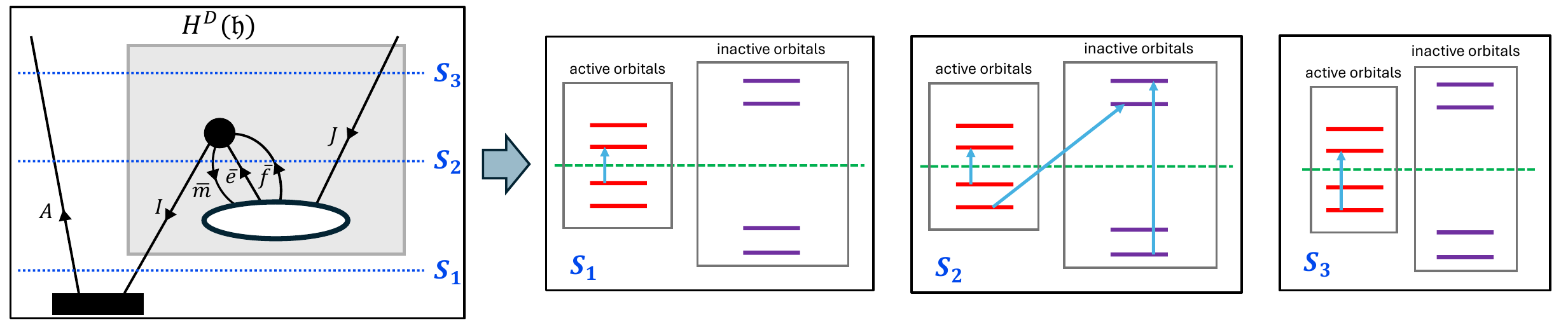}
\caption{
Schematic representation of many-body processes encapsulated in the many-body structure of $H^{\rm D}(\mathfrak{h})$. The leftmost panel represents a one-body Hugenholtz diagram contributing to the effective Hamiltonians $H^{\rm D}(\mathfrak{h})$. The black circle represents the two-body interactions of the bare Hamiltonians, the white oval represents the doubly excited  external cluster amplitude, whereas the black rectangle corresponds to the singly excited Slater determinant
$|\Phi_I^A\rangle = a^{\dagger}_A a_I |\Phi\rangle$ ($a_p^{\dagger}$/$a_p$ represent creation/annihilation operators for electron in $p$-th spin-orbital).
The intermediate states $S_1$, $S_2$, and $S_3$ correspond to the following Slater determinants: $|\Phi_I^A\rangle$, 
$|\Phi_{IJ\bar{m}}^{A\bar{e}\bar{f}}\rangle$ and $|\Phi_J^A\rangle$. The indices $I,J,...$ ($A,B,...$) correspond to the active occupied (unoccupied) spin-orbital indices (occupied/unoccupied in $|\Phi\rangle$), whereas indices $\bar{m},\bar{n}$ ($\bar{e},\bar{f},...$) correspond to the inactive occupied (unoccupied) spin-orbitals. 
The three panels to the right illustrate the electron excitation processes corresponding to states $S_1$, $S_2$, and $S_3$. The green horizontal lines in these panels separate occupied orbitals from unoccupied ones. 
}
\label{fig:heff}
\end{figure*}
\end{widetext}

Since in the effective Hamiltonian (\ref{eq5}) the external degrees of freedom are integrated out, the CC formalism can be viewed as a special example of an embedding procedure (see Refs.\cite{kowalski2023sub}) or renormalization algorithm. It is also instructive to scrutinize the many-body structure of the $H^{\rm D}(\mathfrak{h})$ operator and ensuing properties:
\begin{itemize}
    \item Due to the commutative character of all components defining  $T_{\rm ext}(\mathfrak{h})$ the exact form of the effective Hamiltonian can be constructed. As in the standard SR-CC formulations, the effective Hamiltonian will involve terms that are, at most, fourth powers of cluster operators. 
    \item The effective Hamiltonians can be constructed for each SES and the corresponding active space. In Ref.\cite{kowalski2023sub} the number of SESs defined at the orbital level for the CCSD formalism (CC with singles and doubles of Ref.\cite{purvis82_1910}) amounts to 
      \begin{equation}
        n_o (2^{n_v}-1) + n_v(2^{n_o}-1) - n_o n_v \;,
        \label{eq6}
      \end{equation}
      where $n_o$ and $n_v$ refer to the number of occupied and unoccupied orbitals, respectively. 
    \item The SES active-space problems can be integrated into the flows, where total cluster operator $T_{\rm F}$ is defined through the unique excitation defining internal excitations for active spaces involved in the flow. The so-called Equivalence Theorem, see Ref.\cite{kowalski2021dimensionality}, states that the solution of the flow equations defined in such a way is equivalent to the solution of standard SR-CC equations defined by the operator $T_{\rm F}$ and therefor can be viewed as a "correlated" variant of Aufbau principle (in the sense of adding a sub-system described by its active space).  The SR-CC flow equations have been tested for ground- and excited-state simulations \cite{kowalski2018properties}.
    \item The non-Hermitian CC downfolding and flow equations can be extended to the time-domain \cite{kowalski2020sub,kowalski2021dimensionality}. In Ref.\cite{kowalski2020sub} it has been shown that when the fast-varying part of the wave function defined by the $T_{\rm ext}(\mathfrak{h},t)$ is known then the slow-varying dynamics of the entire system can be described by downfolded Hamiltonian $H^{\rm D}(\mathfrak{h},t)$.
    \item The effective Hamiltonian and its many-body structure provide a theoretical framework to describe a sub-system defined by active space as an open system interacting with the surrounding environment. In Fig.~\ref{fig:heff}, we provided a specific example demonstrating that although the effective Hamiltonian preserves the number of active electrons, the intermediate state (the $S_2$ state) can involve the exchange of electrons between the sub-system and surrounding environment. 
\end{itemize}
In ample numerical tests performed for various levels of CC theory and published in Ref.\cite{kowalski2023sub}, it was demonstrated that using Eq.(\ref{eq5}), one can reproduce exact values of CC calculated using the standard formula given by Eq.(\ref{eq3}) even for active spaces that would not provide a sufficient level of correlation effect in diagonalizing bare electronic Hamiltonians.

The properties of the non-Hermitian CC downfolding offer not only a new way of interpreting SR-CC theory but also provide guiding principles for the design of its Hermitian counterpart.

\subsection{Hermitian CC Downfolding}
The Hermitian CC downfolding is based on the utilization of the DUCC ansatz,
\cite{bauman2019downfolding}
which assumes a similar form as Eq.(3), i.e.,
\begin{equation}
|\Psi\rangle = e^{\sigma_{\rm ext}(\mathfrak{h})}  e^{\sigma_{\rm int}(\mathfrak{h})} |\Phi\rangle \;,
\label{eq8}
\end{equation}
where external and internal general-type anti-Hermitian cluster operators $\sigma_{\rm ext}(\mathfrak{h})$ and $\sigma_{\rm int}(\mathfrak{h})$ are expressed in terms of parameters carrying at least one inactive and no inactive spin-orbital indices, respectively. In general, 
due to the non-commutativity of many-body components defining $\sigma_{\rm ext}(\mathfrak{h})$ and $\sigma_{\rm int}(\mathfrak{h})$ the analysis of the DUCC-based Hermitian downfolding is more challenging. It requires several approximate steps in practical applications. As in the non-Hermitian case, we assume that the active space is defined by sub-algebra $\mathfrak{h}$.

In Ref.\cite{kowalski2020sub}, we performed the analysis of the expansion (\ref{eq8}) using disentangled unitary coupled cluster formalism \cite{evangelista2019exact} in a way that it factorizes into two parts defined products defined by cluster amplitudes (or parameters mentioned earlier) carrying only active spin-orbital indices followed by part defined by cluster amplitudes carrying at least one inactive spin-orbital index. 
Applying the Baker–Campbell–Hausdorff formula multiple times to the first and the second part leads Eq.(\ref{eq8}) where $\sigma_{\rm ext}(\mathfrak{h})$ and $\sigma_{\rm int}(\mathfrak{h})$ are defined by complex and non-terminating commutator expansions. If these expansions exist, then the DUCC ansatz (\ref{eq8}) represents the exact ground-state wave function. It can be shown that if the values of general-type
anti-Hermitian cluster operator (in the sense of the discussion of Ref.\cite{lee2018generalized}) is known, then the exact energy can be reproduced  by 
diagonalizing Hermitian Hamiltonian $H^{\rm D}(\mathfrak{h})$ defined as:
\begin{equation}
    H^{\rm D}(\mathfrak{h}) = (P+Q_{\rm int}(\mathfrak{h}))
    e^{-\sigma_{\rm ext}(\mathfrak{h})} H
    e^{\sigma_{\rm ext}(\mathfrak{h})}
    (P+Q_{\rm int}(\mathfrak{h}))
    \label{eq9}
\end{equation}
using any type of solver. It should be noted that, in analogy to non-Hermitian CC downfolding, only external cluster operators are needed. 

The Hermitian downfolding can also be extended to the flow formalisms (e.g., Quantum Flow (QFlow) formulation of Ref.\cite{kowalski2021dimensionality,kowalski2023quantum} introduced in the context of quantum computing). As in the Hermitian case, the QFlow formalism can be exploited to capture the sparsity of quantum systems.

\section{Numerical methods}
\label{sec:methods}
%
When the $T_{\rm ext}(\mathfrak{h})$ or $\sigma_{\rm ext}(\mathfrak{h})$ operators are known or can be accurately approximated, particularly when static correlation effects can be effectively incorporated by carefully selecting the active space, both Hermitian and non-Hermitian effective Hamiltonians (\ref{eq5}) and (\ref{eq9}) serve as frameworks for devising approximate methods to calculate ground-state energies
(see for example Refs.\cite{kowalski2018properties,bauman2022coupledjcp,shee2024static,feldmann2024complete}). 
In this section, our primary emphasis will be on the Hermitian downfolding and exploration of the usual approximations and computational resources required to determine the approximate representations of the effective Hamiltonians. However, many of these considerations can be easily extended to the Hermitian case. 

\subsection{Approximations for Hermitian downfolded Hamiltonians}

In many applications, the second-quantized representation of the $H^{\rm D}(\mathfrak{h})$ operator
\begin{widetext}
\begin{equation}
H^{\rm D}(\mathfrak{h})= \Gamma^{\rm D}_0(\mathfrak{h}) +
\sum_{PQ} g^P_Q(\mathfrak{h}) a_P^{\dagger} a_Q + \frac{1}{4} \sum_{P,Q,R,S} k^{PQ}_{RS}(\mathfrak{h}) a_P^{\dagger} a_Q^{\dagger} a_S a_R 
+ \frac{1}{36} \sum_{P,Q,R,S,T,U}
l^{PQR}_{STU}(\mathfrak{h}) a_P^{\dagger} a_Q^{\dagger} a_R^{\dagger} a_U a_T a_S + \ldots \;,
\label{eq10}
\end{equation}
\end{widetext}
is required. In the above equations $P$, $Q$, $\ldots$ indices represent general-type active spin-orbitals and  $\Gamma^{\rm D}_0(\mathfrak{h})$ is a scalar.
Compared to the second-quantized form of the bare Hamiltonian $H^{\rm B}(\mathfrak{h})$ in the $\mathfrak{h}$-generated active space
\begin{widetext}
\begin{equation}
H^{\rm B}(\mathfrak{h}) = \Gamma^{\rm B}_0(\mathfrak{h})+ \sum_{PQ} h^P_Q(\mathfrak{h}) a_P^{\dagger} a_Q + \frac{1}{4} \sum_{P,Q,R,S} v^{PQ}_{RS}(\mathfrak{h}) a_P^{\dagger} a_Q^{\dagger} a_S a_R 
\label{eq10b}
\end{equation}
\end{widetext}
the $H^{\rm D}(\mathfrak{h})$ operator contains not only higher-than-pairwise interactions but its matrix representation in the active space is denser than the analogous matrix representation of the $H^{\rm B}(\mathfrak{h})$ operator.

Due to the  non-commuting character of components defining the $\sigma_{\rm ext}(\mathfrak{h})$ operator, one has to rely on the finite-rank commutator expansions, i.e., 
\begin{widetext}
\begin{equation}
H^{\rm D}(\mathfrak{h}) \simeq
(P+Q_{\rm int}(\mathfrak{h}))(H +  \sum_{i=1}^{l}  \frac{1}{i!}[
 \ldots [H,\sigma_{\rm ext}(\mathfrak{h})],\ldots ],\sigma_{\rm ext}(\mathfrak{h})]_i (P+Q_{\rm int}(\mathfrak{h}))\;.
\label{eq11}
\end{equation}
\end{widetext}
where $[\ldots[...]\ldots ]_i$ designates rank-$i$ commutator, to obtain approximate many-body form of the 
$H^{\rm D}(\mathfrak{h})$ operator. 
The length $l$ is not only determined by the accuracy requirements but also by the associated cost. 
In typical applications in quantum chemistry including contributions from single, double, and part of the triple commutators provide a good compromise between the cost and accuracies (see Ref.\cite{bauman2022coupledjcp} for more detailed discussion). 
In Ref.\cite{bauman2019downfolding}, we provided an overview of algebraic techniques 
to derive the second quantized form of Hamiltonian (\ref{eq10}). 
Since the discussed formalisms  are of single-reference type,  the most compressed form of the approximate 
effective Hamiltonians are provided by utilizing particle-hole formalism. 

The second and third classes of approximations are associated with the rank of many-body effects included in Eq.(\ref{eq10}) and the approximation protocol for the
$\sigma_{\rm ext}(\mathfrak{h})$ operator. In all our applications (which will be also used throughout this paper) we include one- and two-body effects defined by 
$g^P_Q$ and $k^{PQ}_{RS}$ tensors whereas 
\begin{equation}
    \sigma_{\rm ext}(\mathfrak{h}) \simeq T_{\rm ext}^{\rm CCSD}(\mathfrak{h}) - T_{\rm ext}^{\rm CCSD}(\mathfrak{h})^{\dagger} \;,
    \label{eq12}
\end{equation}
where $T_{\rm ext}^{\rm CCSD}(\mathfrak{h})$ are external part of the cluster operator $T$ obtained in the CCSD calculations. 

Even the most basic non-trivial approximation entails managing a combinatorial explosion in the quantity of contributing diagrams, equivalent to thousands of Hugenholtz diagrams associated with the incorporation of higher-rank commutators. The manual derivation of all diagrams is not only susceptible to errors but is also excessively time-consuming. To address this issue, we have created a symbolic algebra system named \texttt{SymGen} that can automatically derive tensor expressions corresponding to intricate contractions in Wick's Theorem.

Due to the significant numerical overhead involved in multi-dimensional tensor contractions, the output of the software tool \texttt{SymGen} has been integrated with the Tensor Algebra for Many-body Methods (TAMM) parallel tensor library \cite{mutlu2023tamm} to address the high numerical scaling associated with the above approximation scheme. This scaling increases rapidly with the inclusion of higher many-body effects in the effective Hamiltonian and $\sigma_{\rm ext}(\mathfrak{h})$ operator, not to mention the cost of higher-rank commutators. Therefore, there is a clear need for new computational paradigms to advance effective interaction-based computational models. 

\section{Machine learning for  evaluating downfolded Hamiltonians}
\label{sec:ML}


The high cost associated with the evaluation of the one and two-body interactions in the second quantization representation of the effective Hamiltonian calls for an alternative way to calculate these interactions efficiently.  Because the bare interactions are relatively easy to compute, and the one and two-body interactions appearing in the effective Hamiltonian can be viewed as screened (or dressed) interactions that are related to the bare interactions by a nonlinear map, it may be possible to construct a neural network (NN) representation of such a map denoted by \eqref{eq13} and train such a NN with computed bare interactions and a subset of effective interactions we can afford to compute. 
\begin{equation}
H^{\rm B}(\mathfrak{h})
\xrightarrow{\text{NN}}
H^{\rm D}(\mathfrak{h}) \;.
\label{eq13}
\end{equation}

In particular, we may use bare and effective interactions computed for certain molecular geometries as the training data and use the trained NN to predict effective interactions from bare interactions at other molecular geometries.  We can also view the NN as a computational tool to complete the three- or five-dimensional 
(two- or four-dimensional 
effective interaction tensors with additional dimension corresponding to set of molecular geometries),
from a subset of the interaction tensor elements. 
We will illustrate the performance of NN techniques by using the example of two-body effective interactions. 

\subsection{Physics-Informed Neural Network}
Although it is possible to use a generic deep NN that takes orbital indices such as $P$, $Q$, $R$, $S$ as well as the numerical value of $v^{PQ}_{RS}$ as the input and predicts the value of $k^{PQ}_{RS}$ as the output, such a NN is likely to be less effective than a carefully designed and physics-informed NN~\cite{karniadakis2021physics} 
(for the sake of notational  simplicity, we will sometimes skip the sub-algebra  index $\mathfrak{h}$ in the interaction tensors).
PINNs
have recently proven effective at solving partial differential equations that arise in the physical sciences and engineering applications~\cite{karniadakis2021physics,kissas2020machine}. Specifically, PINNs incorporate physical laws, such as conservation principles or equations of motion, directly into the training process, enhancing the model generalization~\cite{raissi2019physics}. This approach has been successfully applied across various domains, including fluid dynamics, quantum mechanics, and materials science~\cite{raissi2020hidden,norambuena2024physics}.
 Because one and two-body interactions are naturally represented as integrals of molecular orbitals connected by an interaction kernel, the PINN we construct will preserve such a structure.  In particular, we will use a NN to represent screened molecular orbitals and a separate matrix to represent an effective interaction kernel. These components are connected as shown in Figure~\ref{fig:nn} to predict the value of $k^{PQ}_{RS}(\mathfrak{h},\mathbf{R})$ for a given geometry $\mathbf{R}$ (see also the orbital variant $(pq|rs)_D(\mathbf{R})$ of these tensors defined in next subsection). This approach allows us to reduce the number of parameters in the PINN compared to a generic NN. We will describe how such a NN, which we call VNet, is constructed and trained.

It is important to note that a PINN may not be universally applicable to all types of chemical transformations. Instead, we intend to train various NNs based on the nature of the problem to be solved, which is typically determined by the size of the active spaces and the level of approximation used to evaluate effective Hamiltonians. For instance, for a wide range of chemical transformations that involve breaking a single bond, we will utilize NNs trained for relatively small active spaces.
In contrast, for problems involving the breaking of multiple bonds, larger active spaces will necessitate the use of NNs trained for such purposes.


The cost and fidelity of NNs are contingent upon the active space size and rank of the many-body interactions. As a proof of concept, in this paper, we focus on two-body interactions (described by the $k^{PQ}_{RS}(\mathfrak{h})$ tensor) and active spaces, where all occupied orbitals are defined as active.
It also epitomizes the challenging situation for NN algorithms because entries $k^{PQ}_{RS}(\mathfrak{h})$ elements where $P$, $Q$, $R$, and $S$ indices correspond to low-lying orbitals can assume large values. 

We will use the Mulliken-type notation for the orbital representation $(p|q)_B$/$(pq|rs)_B$ and 
$(p|q)_D$/$(pq|rs)_D$
for the one- and two-body interaction tensors of bare and effective Hamiltonians, respectively, where indices $p$, $q$, $r$, $s$ designate the active orbitals. 

The $(p,q,r,s)$th element of the bare two-body interaction tensor is defined by
\begin{equation}
(pq|rs)_B
= \int d\bfr' \int d\bfr \phi_p(\bfr)\phi_q(\bfr) \frac{1}{|\bfr-\bfr'|} \phi_r(\bfr') \phi_s(\bfr'),
\label{eq:vijkl}
\end{equation}
where $\phi_p(\bfr)$ is the $p$th active molecular orbital. When a Gaussian basis set is used to represent $\phi_p$'s, the integral in \eqref{eq:vijkl} can be performed analytically.  A similar integral can be written for the one-body interaction.  For brevity, we will focus on the two-body interaction below.

Our work makes the assumption that the effective two-body interaction for the downfolded Hamiltonian $H^{\rm D}(\mathfrak{h})$ can also be written as a double integral except that the bare 
 Coulomb interaction kernel $1/|\bfr-\bfr'|$ in \eqref{eq:vijkl} should be replaced by a dressed (effective) interaction kernel $W^D(\bfr,\bfr')$, and the active molecular orbitals $\{\phi_p\}$ may need to be modified also to account for change of symmetry in the effective two-body interaction tensor and the nature of the effective interaction.

In general, we do not know the analytical form of $W^D(\bfr,\bfr')$. Although it may be possible to construct a parameterized analytic or even a neural network model for such a kernel and determine the model parameters through a fitting procedure that minimizes the difference between a subset of the effective interaction tensor elements predicted by the model and those computed from the downfolding theory, it is likely to be difficult to evaluate the double integral that defines the effective interaction accurately using such an ansatz.

\subsection{VNet Neural Network }
\begin{widetext}
\begin{figure*}
\includegraphics[scale=0.30]{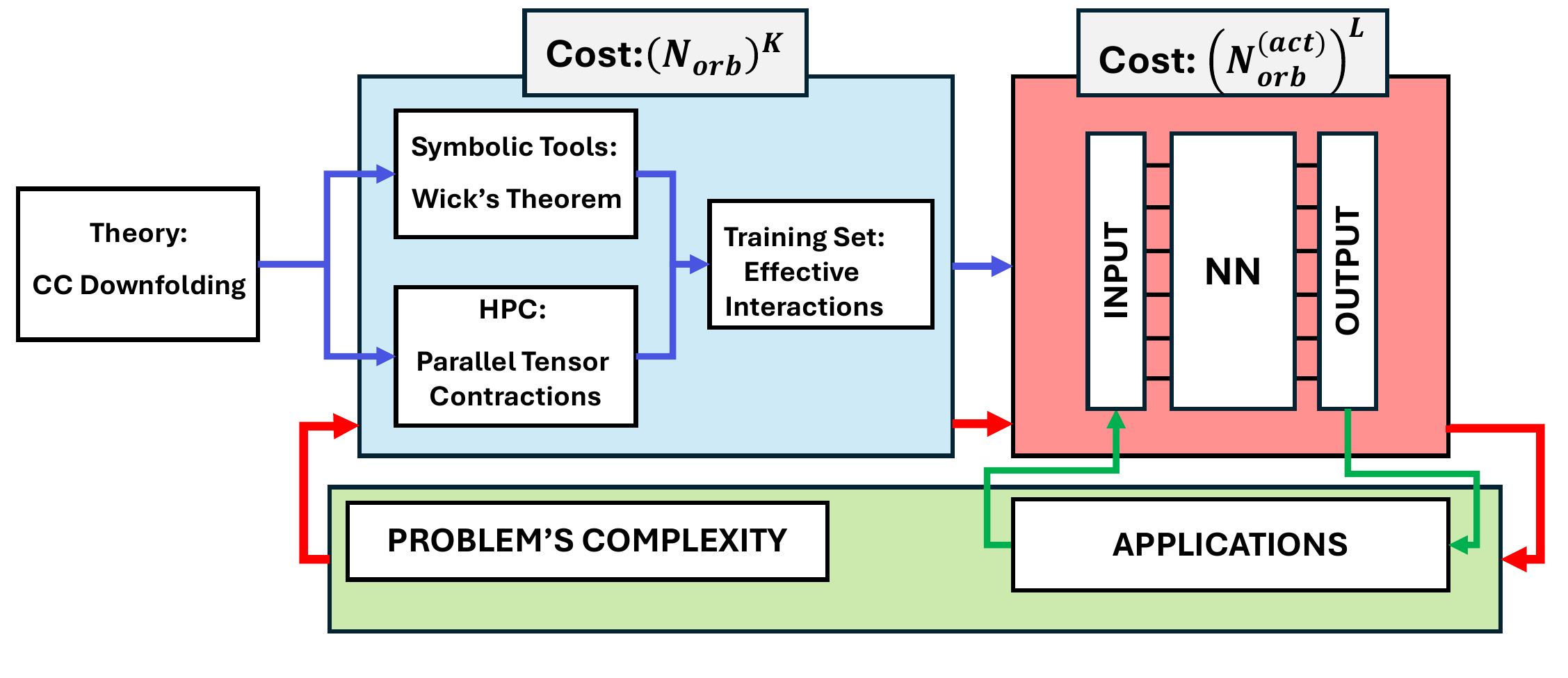}
\caption{
Schematic representation of PINNs for evaluating effective interactions. Training data is aware of symmetries of bare and interaction tensors. Moreover, the PINNs are aware of the problem's complexity, reflected by the size of the active space used. $N_{\rm orb}$ and
$N^{\rm (act)}_{\rm orb}$ designate the total number of orbitals and the number of active orbitals, respectively. It is assumed that $N^{\rm (act)}_{\rm orb}\ll N_{\rm orb}$.
}
\label{fig:nn1}
\end{figure*}
\end{widetext}

\begin{widetext}
\begin{figure*}
\includegraphics[scale=0.55]{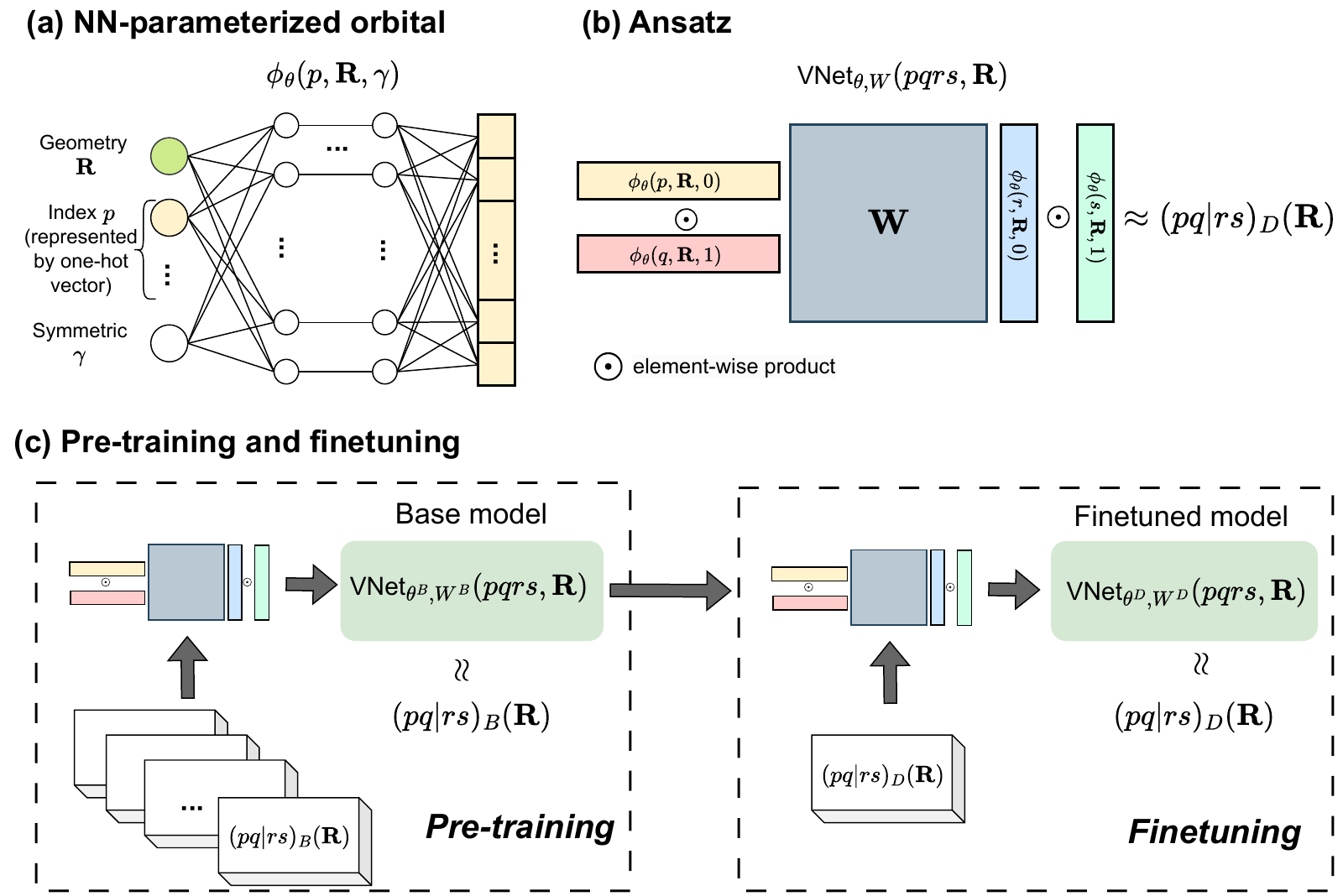}
\caption{\label{Vnet}
VNet. (a) The neural network representation of the molecular orbital $p$. (b) The proposed ansatz to approximate the effective two-body interaction tensor. (c) The proposed paradigm for pre-training and finetuning.
}
\label{fig:nn}
\end{figure*}
\end{widetext}

In this work, we propose an alternative approach schematically depicted in Fig.~\ref{fig:nn1}.  This approach consists of two steps
\begin{enumerate}
\item In the first step, we take a bare two-body interaction tensor $(pq|rs)_B$
and try to decompose it as
\begin{equation}
(pq|rs)_B
= \left[\phi_p\odot\phi_q\right]^T W^B \left[ \phi_r \odot \phi_s\right],
\label{eq:vdecomp}
\end{equation}
where $\phi_p$, $\phi_q$, $\phi_r$, $\phi_s$ are vectors of length $\ell$, chosen from a set of $N_{\rm orb}^{\rm (act)}$ vectors $\{\phi_1,...,\phi_{N_{\rm orb}^{\rm (act)}}\}$, $W^B$ is an $\ell \times \ell$ symmetric matrix, and $\phi_p\odot\phi_q$ denotes the elementwise prodcut of $\phi_p$ and $\phi_q$. 
We perform the decomposition by minimizing the difference between the left and right-hand sides of \eqref{eq:vdecomp} with respect to 
$\{\phi_1,\phi_2,...,\phi_{N_{\rm orb}^{\rm (act)}}\}$, and the unique matrix elements in $W^B$.  The total number of parameters involved in the optimization problem is $N_{\rm orb}^{\rm (act)}\times \ell + (\ell+1)\ell/2$ in this case.  Although we can possibly achieve an accurate fit by making $\ell$ sufficiently large, more flexibility is provided if $\phi$ is represented by a neural network (NN) parameterized by a set of weights and biases contained in a vector $\btheta$. The neural network takes the orbital index as one of the inputs. Because molecular orbitals depend on the geometry of the molecule, the neural network should take the geometry represented by a parameter $\mathbf{R}$ as the input also.  Moreover, we can introduce another parameter $\gamma \in \{0,1\}$ to account for special symmetry of the 
$(pq|rs)_B$
tensor. We will denote this NN by $\phi_{\btheta}(i,\mathbf{R},\gamma)$. The output of the NN contains $\ell$ real numbers.  The architecture of this NN is shown in Figure~\ref{fig:nn}.  
For bare interaction tensors, we always choose $\gamma=0$ so that the 8-fold symmetry of the tensor is preserved. 

We use the bare interaction tensors for several different geometries $\{\mathbf{R}\}$ to train the NN to obtain an initial set of parameters $\btheta$ and the $W^B$ matrix. For each geometry, the training data consists of elements of 
$(pq|rs)_B$
that belong to one nonsymmetric unit. Note that the algebraic decomposition of the bare two-body interaction tensor in \eqref{eq:vdecomp} eliminates physical spatial coordinate $\bfr$ of a molecular orbital and replaces it with a latent spatial coordinate.
\item In the second step, we use the effective two-body interaction tensor elements 
$(pq|rs)_D(\mathbf{R})$
associated with a few selected geometries $\mathbf{R}_i$, $i=1,2,...,n_{\rm ref}$, to refine the NN parameters contained in $\btheta$ and obtain an approximation to the effective interaction kernel $W^D$ by solving the following optimization problem
\begin{widetext}
\begin{equation}
\min_{(W^D)^T = W^D, \btheta}  \sum_{i=1}^{n_{\rm ref}} \sum_{\{p,q,r,s\}\in \mathcal{S}} \biggl | 
(pq|rs)_D(\mathbf{R}_i)
- \left[\phiNN(p,\mathbf{R}_i,0)\odot \phiNN(q,\mathbf{R}_i,1)\right]^T W^D \left[\phiNN(r,\mathbf{R}_i,0)\odot \phiNN(s,\mathbf{R}_i,1) \right]\biggr |^2,
\end{equation}
\end{widetext}
where $\mathcal{S}$ is a subset of the full index set for the $N_{\rm orb}^{\rm (act)}\times N_{\rm orb}^{\rm (act)}\times N_{\rm orb}^{\rm (act)} \times N_{\rm orb}^{\rm (act)}$ tensor 
$(pq|rs)_D(\mathbf{R})$
that belongs to a nonsymmetric subunit, and $n_{\rm ref}$ is the number of selected reference geometries.  Note that  $(pq|rs)_D$ only has a 4-fold symmetry. 
Calculating $(pq|rs)_D$, in contrast to $(pq|rs)_B$,  is associated with a significant numerical overhead.

It is important to point out that both 
$(pq|rs)_B(\mathbf{R})$
and 
$(pq|rs)_D(\mathbf{R})$
are geometry dependent. This $\mathbf{R}$ dependency translates into the dependence of $\phiNN$ on $\mathbf{R}$.  However, both the bare and effective interaction kernels $W^B$ and $W^D$ are independent of $\mathbf{R}$. As a result, we can choose $\mathcal{S}$ to be the entire index subset for a nonsymmetric subunit of 
$(pq|rs)_D(\mathbf{R})$
and determine $W^D$ from the effective two-body interaction tensor for a few geometric configurations associated with a few different $\mathbf{R}_i$'s.  This $W^D$ can be combined with the refined NN $\phiNN(i,\mathbf{R}',\gamma)$ to predict effective two-body interaction tensors at other geometries defined by a different set of $\mathbf{R}'$.
\end{enumerate} 
%
%
%
%
%
%
In the forthcoming analysis, we aim to present a proof of concept for the applicability of the proposed algorithms in small-sized benchmark systems, namely H$_2$O and HF molecules. Our approach will showcase the potential benefits of training neural networks (NNs) using datasets acquired from the immediate vicinity of equilibrium geometries, which is motivated by the fact that SR-CC formalisms yield the most accurate results in such regions. Subsequently, we will employ the NN-predicted effective Hamiltonians to explore the other parts of the configurational space of the molecular systems, for example, those related to the stretching or breaking of a single bond.

\section{Results and discussions}
\label{sec:results}
We now demonstrate the effectiveness using the bare and effective two-body interactions of two molecules H$_2$O and HF at a few selected reference geometries to train VNet and predict (complete) the two-body effective interactions of these molecules at other geometries. For both H$_{2}$O and HF, we first computed the complete active space self-consistent field (CASSCF) reference with the active space consisting of the eight lowest-energy orbitals and the cc-pVTZ basis set. 
The final effective interactions were then computed using the A7 DUCC approximation that is described in Ref.~\cite{bauman2022coupledjcp}.

For the H$_2$O molecule, we used the bare two-body interaction tensors associated with different relative O-H bond lengths corresponding to a single bond breaking, i.e., the ratio between the O-H bond length ($R_{\rm OH}$) and the O-H bond length at equilibrium ($R_{\rm eq}$; $ R_{\rm eq} = {\rm 1.84345\; a.u.}$), ranging from 1.1 to 2.5 (a total of 66 different geometries) to train the VNet to obtain both $W^B$ and an initial set of parameters $\theta$ for $\phiNN$. We used the effective two-body interaction tensors associated with relative bond lengths 1.15, 1.45, 1.95, and 2.45 to finetune the parameter $\theta$ in the NN representations of the molecular orbitals $\phiNN$ and to obtain an effective interaction kernel matrix $W^D$. The NN used to represent $\phiNN$ consists of 
$N_{\rm orb}^{\rm (act)}+2$
input neurons. This includes 
$N_{\rm orb}^{\rm (act)}$
neurons in a one-hot encoding to account for the orbital index, one neuron for the geometry $\mathbf{R}$, and one neuron for the symmetry index $\gamma$.  The NN contains 3 hidden layers with 200 neurons in each layer. The output of the NN consists of 300 neurons and we represent $W^B$ and $W^D$ as $300\times 300$ symmetric matrices. The implementation details, such as optimization specifics, are provided in Appendix~\ref{app:implement}. 

In Figure~\ref{fig:H2Opred}, we plot the maximum absolute error (MAE) and mean square error (MSE) of the two-body effective interaction tensor elements for several geometries associated with different relative O-H bond lengths (orange curve marked with dots). We observe that the MAEs for all geometries are around $10^{-3}$, which are sufficiently small. Not surprisingly, the MAEs of the predicted effective tensors associated with geometries close to the geometries used for training (marked by green crosses) are much smaller. As a comparison, we also plot the maximum difference between an element of the bare two-body interaction tensor and the corresponding element in the effective two-body tensor.  We can clearly see this difference is almost two orders of magnitude larger.

\begin{figure}
    \centering
    \includegraphics[width=0.49\textwidth, trim = {1cm 0 0 0}, clip]{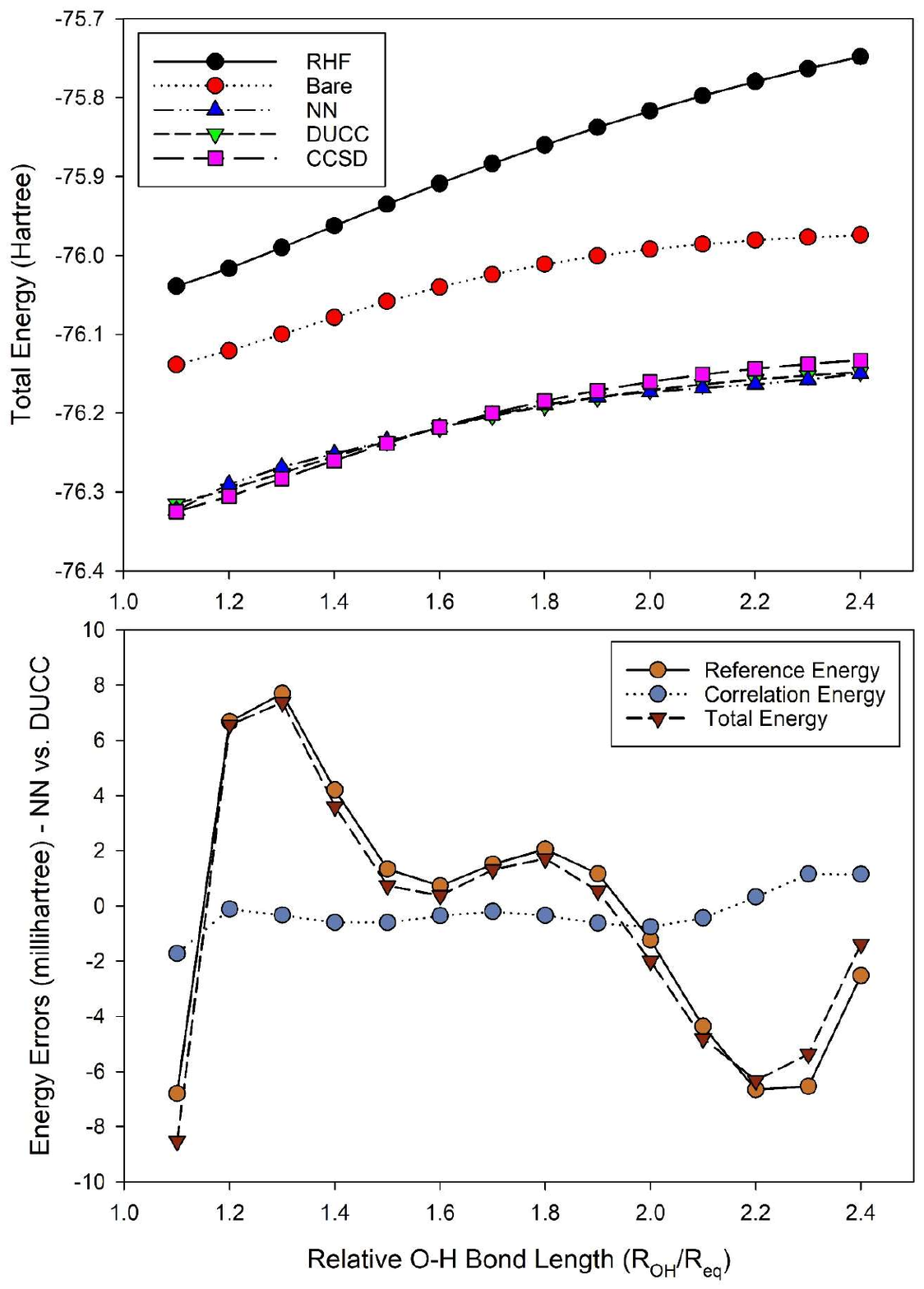}
    \caption{Top Panel: Comparison of total energies of H$_2$O obtained from the RHF, DUCC, and neural network Hamiltonians. Bottom Panel: Analysis of the energy components between the neural network Hamiltonians and corresponding DUCC Hamiltonians. The x-axis is the $\rm O-H$ bond distance for the stretched bond along the single $\rm O-H$ bond breaking pathway and is measured relative to the equilibrium bond length (${\rm R}_{\rm eq} = {\rm 1.84345 a.u.}, {\rm H-O-H} = 110.6^{\circ}$).} 
    \label{fig:H2O-Energies}
\end{figure}

To further evaluate the accuracy of the neural network-derived Hamiltonians, we computed the reference, correlation, and total energies for the Hamiltonians and compared them against the expected DUCC Hamiltonians. The one-body contribution to the Hamiltonian was fixed to the corresponding Hamiltonian (either bare or downfolded). As seen in Fig.~\ref{fig:H2O-Energies}, the DUCC Hamiltonians well capture the correlation along the bond-breaking coordinate and are a significant improvement over a simple truncation of the Hamiltonian to the active space, which is illustrated by the bare results. For all geometries, the neural network generated Hamiltonians provide energies close to the expected DUCC results. In fact, the neural network Hamiltonians capture over 97\% of the correlation energy for all geometries. 
The largest source of error in the total energies is attributed to the two-body interactions of the largest magnitude, which are represented by the reference energy ($(ii|jj)$ and $(ij|ji)$, where $i$ and $j$ are occupied spin-orbital indices). For these large-magnitude interactions, even small relative errors can lead to measurable errors in energies. We plan to investigate this further in later studies to find approaches to better address this class of interactions.

For the HF molecule,
we used the bare two-body interaction tensors associated with different relative H-F bond lengths ($R_{\rm HF}/R_{\rm eq}$; $R_{\rm eq} = {\rm 1.7325 a.u.}$) ranging from 0.85 to 2.0 (a total of 215 different geometries) to train the VNet to obtain both $W^B$ and an initial set of parameters $\theta$ for $\phiNN$. We used the effective two-body interaction tensors associated with relative bond lengths 0.95, 1.35, 1.65, and 1.95 to finetune the parameter $\theta$ in the NN representations of the molecular orbitals $\phiNN$ and to obtain an effective interaction kernel matrix $W^D$. The NN architecture remains the same for HF.

\begin{figure*}
\includegraphics[width=0.95\textwidth]{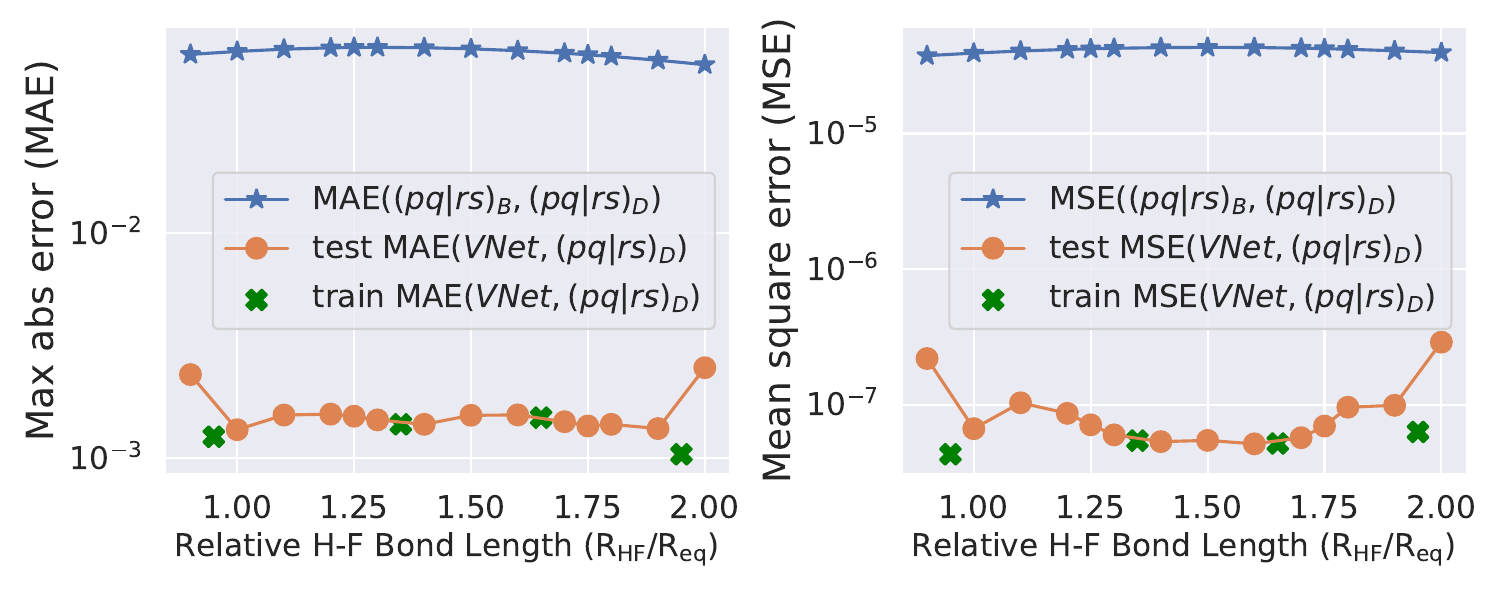}
\caption{
Maximum absolute error (MAE) and mean square error (MSE) between the true two-body effective interaction tensor and the predicted tensor for the HF molecule.
}
\label{fig:HFpred}
\end{figure*}

In Figure~\ref{fig:HFpred}, we plot the MAE and MSE of the two-body effective interaction tensor elements for several geometries associated with different relative H-F bond lengths (orange curve marked with dots). We observe that the MAEs for all geometries are around $10^{-3}$, which are sufficiently small. Not surprisingly, the MAEs of the predicted effective tensors associated with geometries close to the geometries used for training (marked by green crosses) are smaller. As a comparison, we also plot the maximum difference between an element of the bare two-body interaction tensor and the corresponding element in the effective two-body tensor.  We can clearly see this difference is almost two orders of magnitude larger.

\begin{figure}
    \centering
    \includegraphics[width=0.49\textwidth, trim = {0.8cm 0 0 0}, clip]{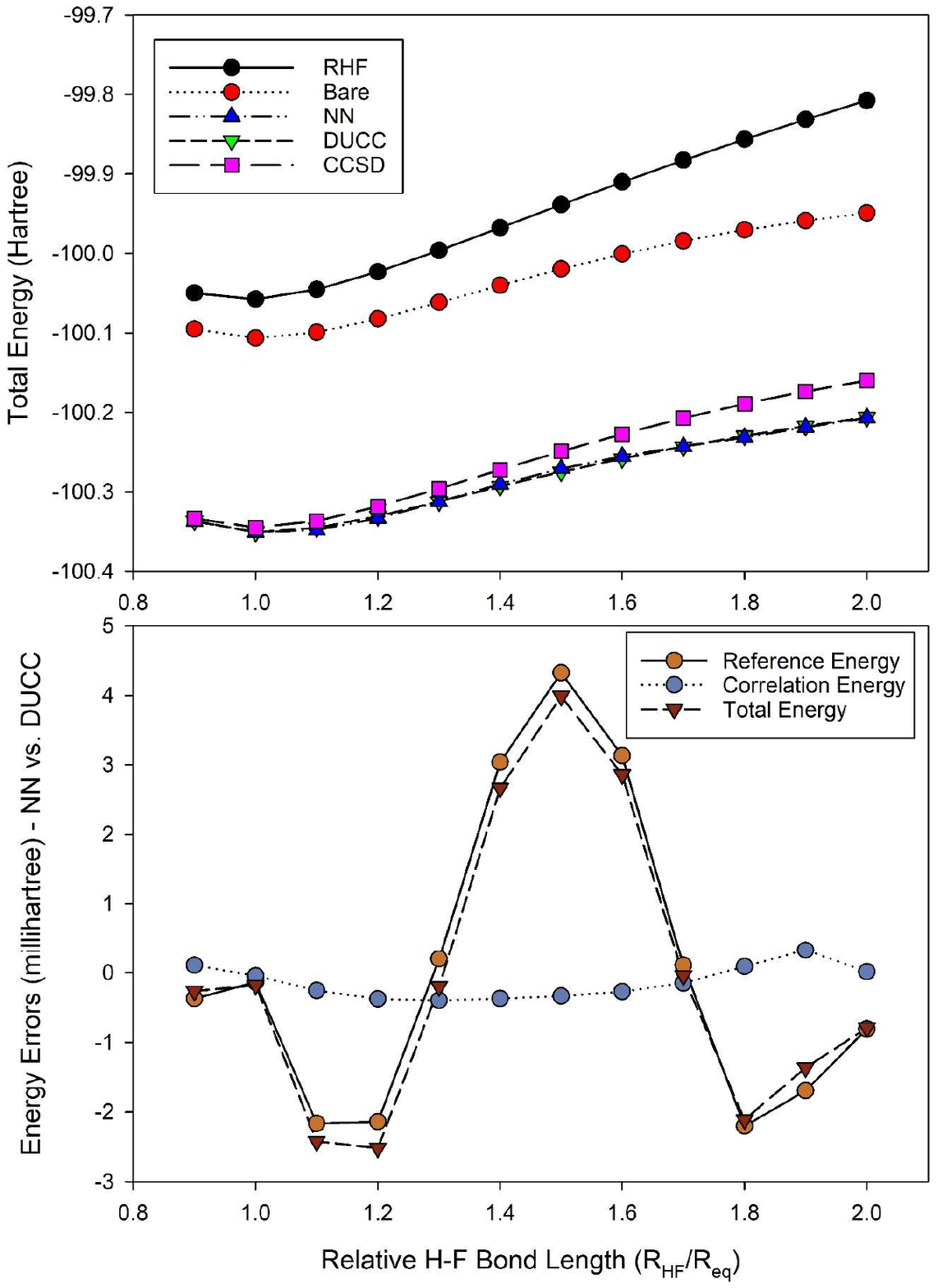}
    \caption{Top Panel: Comparison of total energies of HF obtained from the RHF, DUCC, and neural network Hamiltonians. Bottom Panel: Analysis of the energy components between the neural network Hamiltonians and corresponding DUCC Hamiltonians. The x-axis is the H--F bond distance relative to the equilibrium bond length (${\rm R}_{\rm eq} = {\rm 1.7325 a.u.}$).} 
    \label{fig:HF-Energies}
\end{figure}

As with H$_2$O, we computed the reference, correlation, and total energies for the Hamiltonians of HF and compared them against the expected DUCC Hamiltonians to evaluate the accuracy of the neural network-derived Hamiltonians. 
Fig.~\ref{fig:HF-Energies} shows that the DUCC Hamiltonians well capture the correlation and are a significant improvement over the truncated bare Hamiltonian in the active space once again. It is important to emphasize that any error in DUCC can be systematically improved through the underlying approximations as shown in Ref.~\onlinecite{KowalskiDUCCAccuracies}. For all geometries, the neural network generated Hamiltonians provide energies close to the expected DUCC results capturing over 99\% of the correlation energy for all geometries. Again, the largest source of error in the total energies is attributed to the two-body interactions of the largest magnitude which are represented by the reference energy. Compared with H$_2$O, both the reference and correlation energies better agree with the expected DUCC results in general.

It takes $(1.80\pm0.12)\times 10^{-3}$ seconds (averaged over 10 independent trials) to predict the effective interaction for H$_2$O in a particular geometry using the VNet. This is a drastic saving over the relatively costly downfolding procedure which has $\mathcal{O}(\mathscr{N}^6)$ scaling with the system size $\mathscr{N}$ for the underlying CCSD calculation and similar scaling for the actual downfolding procedure.

\begin{figure*}[ht]
\centering
	\subfigure[The HF molecule]{\includegraphics[width=0.48\linewidth ]{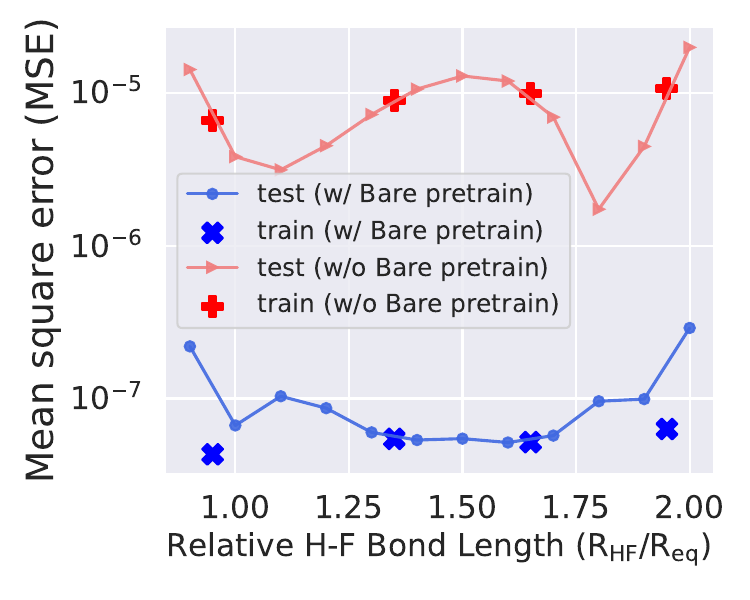}}
	\subfigure[The H$_2$O molecule]{\includegraphics[width=0.48\linewidth ]{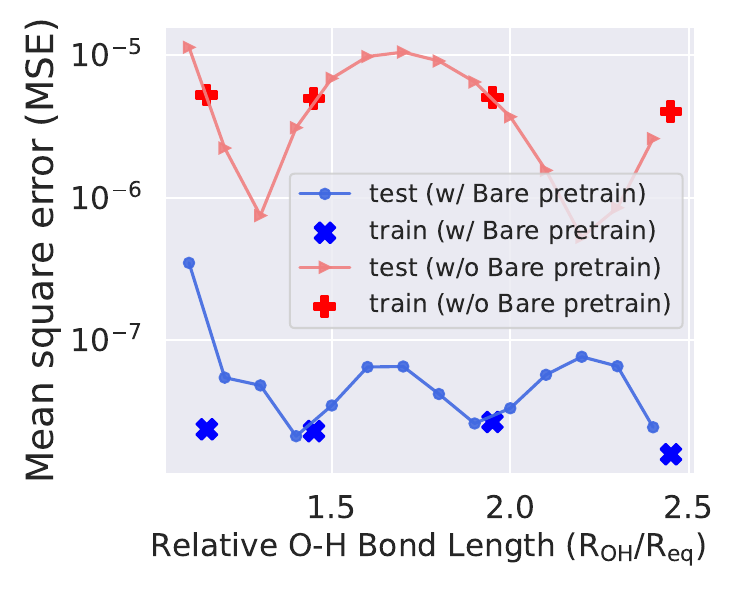}}
\caption{Comparison of NN training on DUCC with and without bare pretrained orbitals on the molecules (a) HF and (b) H$_2$O.}
\label{fig:pretrain}
\end{figure*}

Figure~\ref{fig:pretrain} illustrates the necessity to use the bare interaction to  pretrain VNet to obtain a good initial guess of the VNet parameters. Without using the bare interaction to pretrain VNet, the MSE of the VNet loss function plateaus around $10^{-5}$. In contrast,  after using the bare interation to pretrain the VNet first and then using the DUCC effective interaction to finetune to VNet parameters, the MSE of the loss function reaches $10^{-7}$, which is two orders of magnitude lower than the non-pretrained version. 

\begin{figure*}[ht]
\centering
	\subfigure[The HF molecule]{\includegraphics[width=0.48\linewidth ]{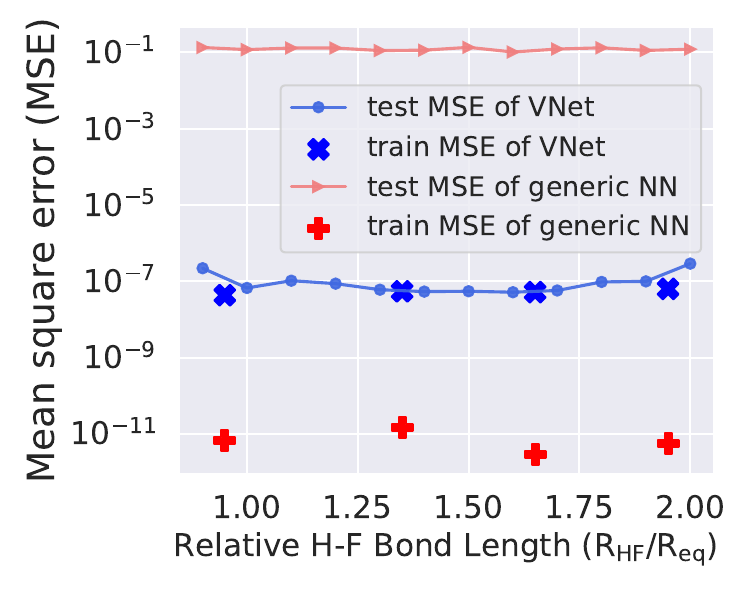}}
 \subfigure[The H$_2$O molecule]{\includegraphics[width=0.48\linewidth ]{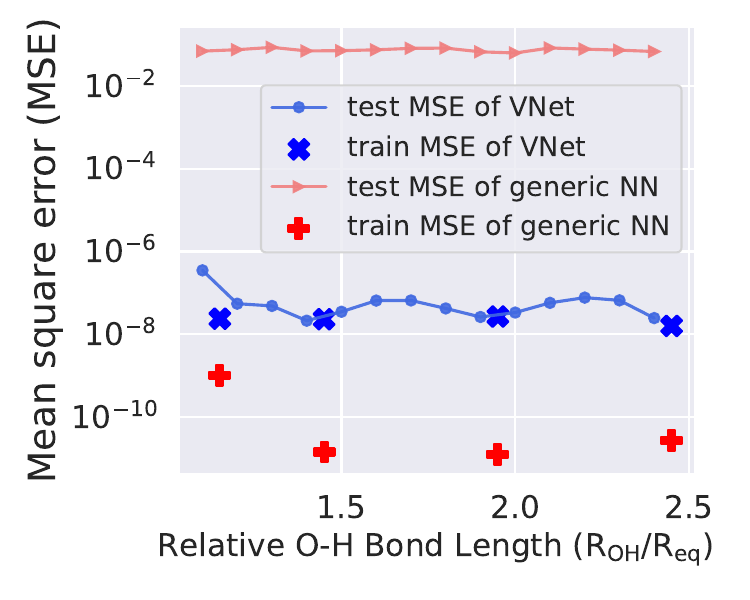}}
\caption{Comparison of training and testing performance for regression using Vnet and Coordinate-based MLP methods on the molecules (a) HF and (b) H$_2$O. The Fourier feature is a coordinate-based MLP approach that uses a special fully connected network with input coordinates and outputs the corresponding value. Although the training error of the Fourier feature method is nearly zero, the testing error is relatively large, indicating a severe overfitting problem.}
\label{fig:fourier}
\end{figure*}

In Figure~\ref{fig:fourier}, we compare the effectiveness of VNet and that of a generic fully connected neural network with Fourier features~\cite{tancik2020fourier} that takes the indices of 
$(pq|rs)_D$
as well as the geometry $\mathbf{R}$ as the input and predicts the numerical value of the corresponding tensor element as the output.  The generic NN with Fourier features is trained with the effective interaction tensor at a few selected geometries. The details are provided in Appendix~\ref{app:coordinatemLP}. As we can see from this figure, the training mean square error (MSE) can be as low as $10^{-11}$. However, when this NN is used to infer effective interaction tensor elements outside of the training geometries, the MSE  can be as large as $10^{-1}$. Clearly, the generic NN, which does not take advantage of the special structure of the problem, is overfitted to the training data and has trouble completing effective interaction tensors for untrained geometries.

In addition to producing accurate approximations to effective interaction tensors for different geometries, VNet also allows us to study the effect of screening by comparing the bare and effective interaction kernel matrices $W^B$ and $W^D$ which are independent of the geometry.

We computed the eigendcomposition of $W^B$, i.e.,
\begin{equation}
W^B = Z \Lambda^B Z^T,
\end{equation}
where the matrix $Z$ contains the eigenvectors of $W^B$, and the diagonal matrix $\Lambda^B$ contains the corresponding eigenvalues $\varepsilon_1^B,\varepsilon_2^B,...,\varepsilon_{\ell}^B$.

\begin{figure*}[ht]
\centering
	\subfigure[$Z^T W^D Z$]{\includegraphics[width=0.48\linewidth ]{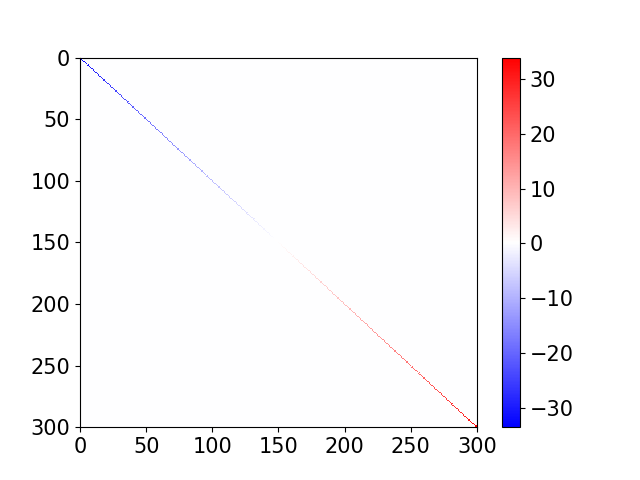}}
	\subfigure[$(\varepsilon^D - \varepsilon^B)/\varepsilon^B$]{\includegraphics[width=0.40\linewidth ]{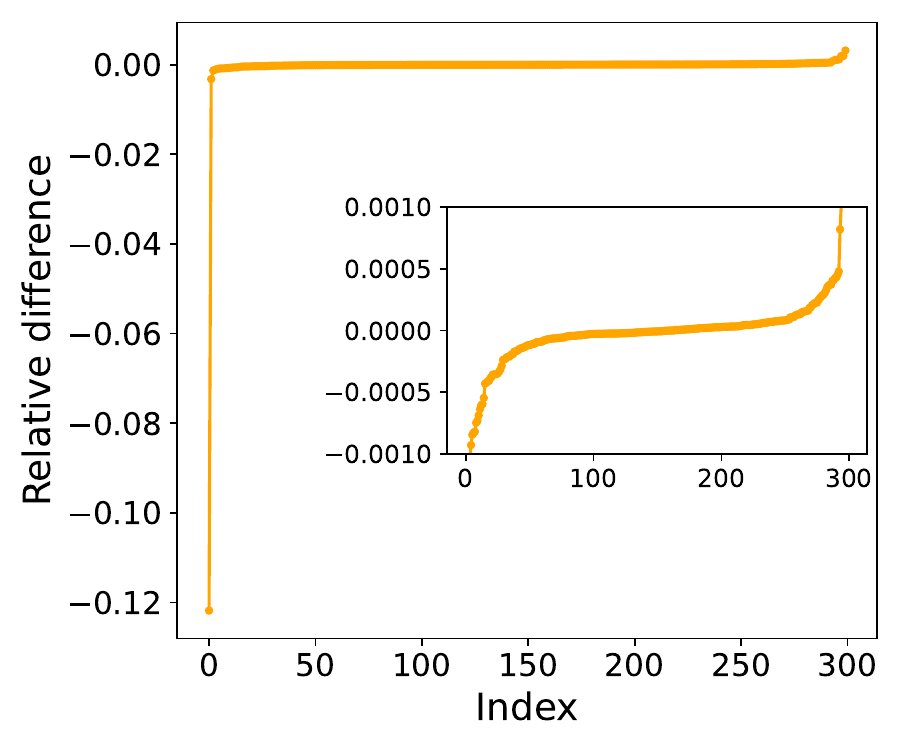}}
\caption{Analyzing the bare and effective interaction kernel matrices $W^B$ and $W^D$ for the H$_2$O molecule. (a) $Z^T W^D Z$, where $Z$ denotes the eigenvectors of $W^B$. (b) The ratio of all eigenvalues of $W^D$ to the corresponding eigenvalues of $W^B$. }
\label{fig:screening}
\end{figure*}

\begin{figure*}[ht]
\centering
	\subfigure[$Z^T W^D Z$]{\includegraphics[width=0.48\linewidth ]{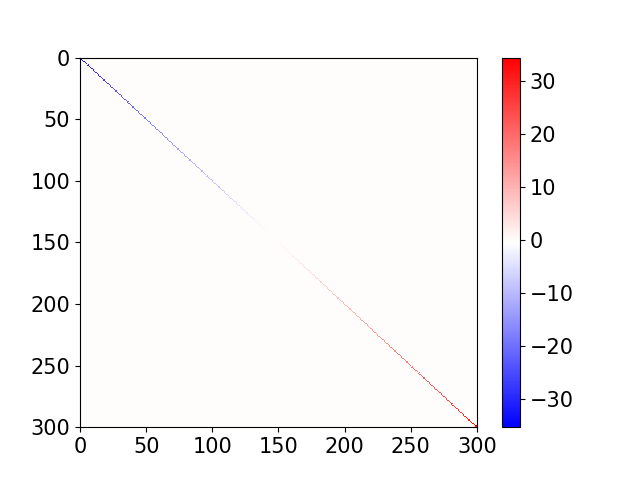}}
	\subfigure[$(\varepsilon^D - \varepsilon^B)/\varepsilon^B$]{\includegraphics[width=0.40\linewidth ]{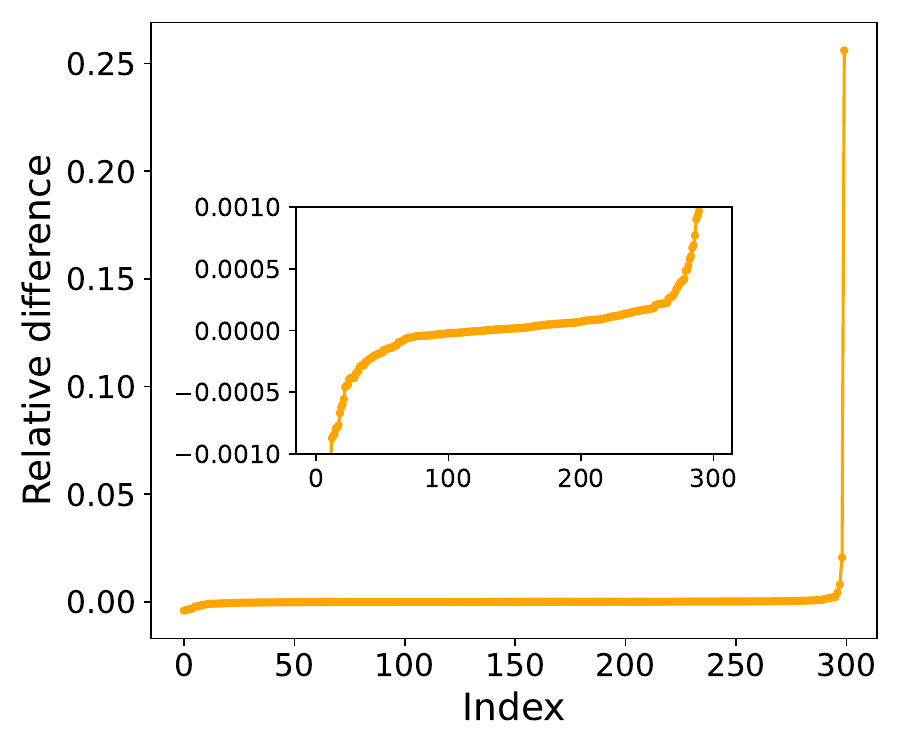}}
\caption{Analyzing the bare and effective interaction kernel matrices $W^B$ and $W^D$ for the HF molecule. (a) $Z^T W^D Z$, where $Z$ denotes the eigenvectors of $W^B$. (b) The ratio of all eigenvalues of $W^D$ to the corresponding eigenvalues of $W^B$.}
\label{fig:screening2}
\end{figure*}

Using the eigenvector matrix $Z$ associated with $W^B$, we compute $Z^T W^D Z$. Figures~\ref{fig:screening} and~\ref{fig:screening2} show that this matrix is nearly diagonal for both H$_2$O and HF, which means that $W^D$ and $W^B$ share the almost same set of eigenvectors.  The diagonal elements of $Z^T W^D Z$ are close to the eigenvalues of $W^D$. Therefore, it is meaningful to compare the eigenvalues of $W^B$ and $W^D$ in order to understand the screening function of the effective interaction kernel.

In Figures~\ref{fig:screening} and~\ref{fig:screening2}, we plotted the ratio $\tau = (\varepsilon^D - \varepsilon^B)/\varepsilon^B$ for all eigenvalues of $W^D$ and the corresponding eigenvalues of $W^B$.  These ratios are sorted in increasing order before they are plotted. Note that the horizontal axis does not correspond to a physical spatial variable.  It is a latent dimension created in the VNet representation of the molecular orbital. 
The inset plots were obtained by excluding data points that correspond to nearly zero $\varepsilon^B$ values. From these plots, we can see that the ratio $\tau$ seems to follow a scaled tangent which suggests that $\varepsilon^D$ is related to $\varepsilon^B$ by
$\varepsilon^D_i = \varepsilon^B_i (1+\beta \tan(\alpha (i-i_c)))$ for some parameters $\alpha$, $\beta$ and $i_c$.




\begin{figure*}
\includegraphics[width=0.95\textwidth]{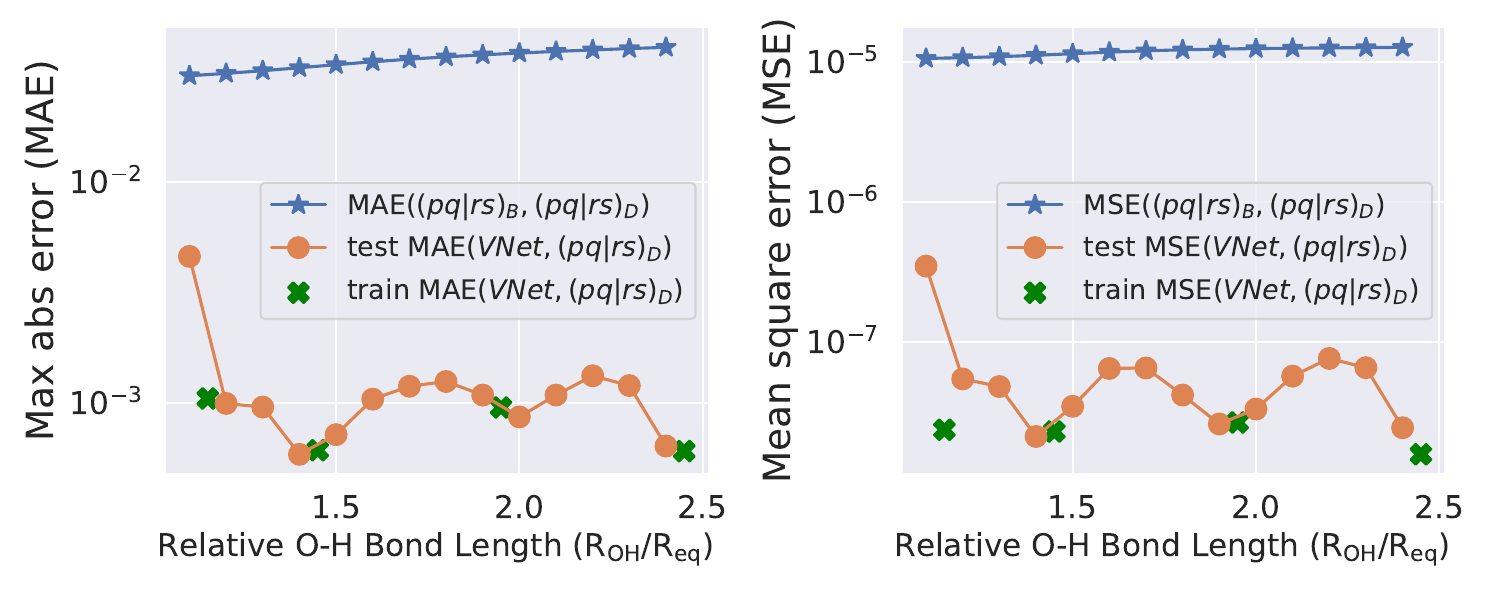}
\caption{Maximum absolute error (MAE) and mean square error (MSE) between the true two-body effective interaction tensor and the predicted tensor for the H$_2$O molecule.
}
\label{fig:H2Opred}
\end{figure*}

\section{Conclusion}
\label{sec:conclusion}
The CC downfolding theory opens the door to studying
large molecules by constructing an effective Hamiltonian
in a small-dimensionality active space.
Although it is possible to construct such a Hamiltonian for a few selected geometries, performing such calculations for a large number of geometries as would required in a dynamics simulation is costly.  In this work, we propose an ML-based approach to predict the effective interactions of the downfolded Hamiltonian by a special physics-informed neural network called VNet. The VNet is trained with several bare interaction tensors as well as a few effective interaction tensors associated with a few selected geometries. Using HF and H$_2$O molecules as examples, we have demonstrated that VNet can be effectively trained to accurately predict the effective interactions for many geometries not included in the training dataset. 
For these problems, the VNet model has 188,050 trainable parameters. For the HF case, training on the 215 Bare tensors for 5,000 epochs takes 4.4 hours. Finetuning VNet using 4 DUCC tensors for 500 epochs takes 10 minutes.  Once VNet is trained, it only takes $(1.80\pm0.12)\times 10^{-3}$ seconds (averaged over 10 independent trials) to predict one effective interaction tensor. This presents a powerful alternative to combat the $\mathcal{O}(\mathscr{N}^6)$ scaling associated with the downfolding procedure.

Not only can VNet allow us to accurately predict effective interactions rapidly, but it can also be used to elucidate the nature of effective interactions in terms of the screening effect of the effective interaction kernel.
In addition, the efficacy of the VNet framework was illustrated in a complex scenario involving active orbitals, comprising a subset of valence virtual orbitals and all occupied orbitals. The latter group of orbitals exhibits diverse localization patterns and corresponds to different energy regimes. For future applications, we intend to adopt a more uniform definition of active orbitals, encompassing only valence occupied and virtual orbitals.

What we have presented here is the first attempt to use PINN-based ML techniques to model and predict effective interactions for realistic molecules. The VNet architecture we used in this work is by no means optimal and can be further improved.  More work is also required to investigate the transferability of this approach to other molecular systems.

\begin{acknowledgments}
This material is based upon work supported by the ``Transferring exascale computational chemistry to cloud computing environment and emerging hardware technologies (TEC$^4$)''  project, which is funded by the U.S. Department of Energy, Office of Science, Office of Basic Energy Sciences, the Division of Chemical Sciences, Geosciences, and Biosciences (under FWP 82037). N.P.B. also acknowledges support from the Laboratory Directed Research and Development (LDRD) Program at Pacific Northwest National Laboratory. 
This work is also supported by the U.S. Department of Energy, Office of Science, Office of Advanced Scientific Computing Research and Office of Basic Energy Science, Scientific Discovery through Advanced Computing (SciDAC) program under Contract No. DE-AC02-05CH11231. (S.L. and C.Y.) This work used the computational resources of the National Energy Research Scientific Computing (NERSC) center under NERSC Award ASCR-ERCAP m1027 for 2024, which is supported by the Office of Science of the U.S. Department of Energy under Contract No. DE-AC02-05CH11231.

\end{acknowledgments}

\appendix

\section{Implementation details}
\label{app:implement}
This section will outline the implementation details, including dataset preparation, neural network architecture, and neural network optimization.

\subsection{Dataset} For the H$_2$O molecule case, bare tensors with geometry ranging from relative H-O bond lengths of 1.1 to 2.5 (a total of 66 different bond lengths)
were used to train a baseline VNet model. Then DUCC effective two-body interaction tensors with relative H-O bond lengths 1.15, 1.45, 1.95 and 2.45 were used to finetune the baseline VNet model. The finetuned model was then used to predict the DUCC effective two-body interaction tensors associated with relative H-O bond lengths ranging from 1.1 to 2.5 in equally spaced increments of 0.1.

For the HF molecule case, bare tensors with geometry ranging from relative H-F bond lengths of 0.85 to 2.0 (a total of 215 different bond lengths) were used to train the baseline VNet model. Then DUCC tensors with relative H-F bond lengths 0.95, 1.35, 1.65, and 1.95 were used to finetune the VNet. The model was then tested to predict the DUCC tensors with relative H-F bond lengths ranging from 0.9 to 2.0 in equally spaced increments of 0.1.

\subsection{Index set for reconstructing DUCC tensor} The size of the index set for a two-body interaction tensor (either DUCC or Bare) is 
$(N_{\rm orb}^{\rm (act)})^4$.
Since the DUCC tensor is 4-fold symmetric, the total number of unique indices is approximately 
$(N_{\rm orb}^{\rm (act)})^4/4$. 
We denote this unique index set by $\mathbb{S}$. A considerable number of two-body interaction tensor elements are nearly zero and these elements do not contribute significantly to the total energy calculation. As a result, we only consider reconstructing the non-zero components of the DUCC tensor. To identify the indices of the zero tensor elements, we use the assumption that $(pq|rs)_D(\mathbf{R}) = 0$ if and only if $(pq|rs)_B(\mathbf{R}) = 0$.

In practice, we set $(pq|rs)_D(\mathbf{R})$ to zero if $\mathrm{mean}\{(pq|rs)_B(\mathbf{R}_i)\} < \epsilon_1$ and $\mathrm{std}\{(pq|rs)_B(\mathbf{R}_i)\} < \epsilon_2$, where the mean and the standard deviation (std) are taken over all geometries indexed by $i$. We denote this set of indices $(p,q,r,s)$ associated with the zero DUCC tensor elements by $\mathbb{P}$. We use $\epsilon_1 = \epsilon_2 = 10^{-5}$ for all tests performed in this work. The index set $\mathbb{P}^C = \mathbb{S}\setminus\mathbb{P}$ represent all nonzero tensor elements we construct. The mean square error (MSE) and maximum absolute error (MAE) are calculated only on $\mathbb{P}^C$. 

\subsection{Network architecture} We approximate molecular orbital by a fully connected neural network, which can be expressed as $f(x)=W_L\sigma(\cdots (W_2\sigma(W_1x+b_1)+b_2))+b_L$. Here, the input $x$ is a vector of length 
$N_{\rm orb}^{\rm (act)}+2$, consisting of three components. The first component is a scalar representing the geometry (bond length) of the molecule. The second component is an integer that describes the type of symmetry of the tensor. The third component is a one-hot vector $h$ of length 
$N_{\rm orb}^{\rm (act)}$.
It is an occupation representation of the orbital index $i\in \{1,2,\cdots,N_{\rm orb}^{\rm (act)}\}$, i.e., $h(i)=0$ and $h(j)=0$ if $j\neq i$. The weight matrices are defined as $W_1\in \mathbb{R}^{(N_{\rm orb}^{\rm (act)}+2)\times M}$, $W_L\in \mathbb{R}^{M\times k}$, and $W_s\in \mathbb{R}^{M\times M}$ for $s=2,\cdots,L-1$. The bias vectors are $b_s$ of length $M$ for $s=1,\cdots,L-1$ and $b_L=k$. The function $\sigma$ is a non-linear activation function applied element-wise to each entry of a vector.

We use the Sigmoid Linear Unit~\cite{hendrycks2016gaussian} as the activation function, i.e., $\sigma(x) = x/(1+\exp(-x))$. The values of $k$, $M$, and $L$ are chosen as 300, 200, and 4, respectively.

\subsection{Optimization} During the training phase, the VNet parameters are optimized using the Adam optimizer, a variant of the stochastic gradient descent method, for 5000 epochs. The learning rate is initialized at 0.001 and follows a cosine decay schedule. In the finetuning phase, these parameters are further optimized for 500 epochs. The learning rate is set to 0.0002 for the H$_2$O molecule case (0.0005 for the HF case), which is smaller than the rate used in the initial training phase. The learning rate is also adjusted according to a cosine decay schedule.
\section{Comparison with generic NN method}
\label{app:coordinatemLP}
To demonstrate the advantage of a PINN over a generic NN for completing the effective two-body interaction, we compare the proposed VNet with a coordinate-based multi-layer perceptron (MLP)~\cite{siren,liang2024,tancik2020fourier}. A coordinate-based MLP is a fully connected neural network architecture often designed to learn continuous functions in several spatial coordinates. Unlike traditional MLPs that operate on abstract data representations, coordinate-based MLPs take a set of coordinates as the input and produce a scalar or vector as the output, approximating the function value, e.g., color or density, defined at this set of coordinates. The Fourier feature method (FFM)~\cite{tancik2020fourier}, a popular coordinate-based MLP that adds a Fourier feature with random frequencies to the input layer, is adopted for comparison in the experiments. The input to the FFM is the index $(p, q, r, s)$ of a particular two-body interaction tensor element as well as the geometry $\mathbf{R}$. The output of FFM approximates $(pq|rs)_D(\mathbf{R})$. We used the same training setup as that used for VNet except that the number of training epochs was set to 200. In Figure \ref{fig:fourier}, we compare the training and testing errors of VNet with those of the coordinate-based MLP methods for both H$_2$O and HF molecules. While the coordinate-based MLP achieves almost zero loss function values on the training data, the testing error on other geometries is relatively large. For VNet, the loss function achieves nearly the same values on both the training and testing datasets. Therefore, the physics-informed model provided by VNet appears to be superior to a generic coordinate-based MLP NN.

\providecommand{\noopsort}[1]{}\providecommand{\singleletter}[1]{#1}%

\end{document}